\documentclass[twocolumn]{aastex62}
\PassOptionsToPackage{table}{xcolor}
\usepackage[table]{xcolor}
\usepackage{tabularx,colortbl}
\usepackage{graphicx}
\usepackage{color,colortbl}
\usepackage{amsmath}
\usepackage{lineno}
\usepackage{multirow}

\usepackage{enumitem}
\usepackage{graphicx}
\usepackage{dcolumn}
\usepackage{bm}
\usepackage{hyperref}
\usepackage{enumitem}

\usepackage{graphicx}
\usepackage{color}
\usepackage{amsmath}
\usepackage{lineno}
\usepackage{multirow}
\usepackage{longtable}

\definecolor{GRAY}{rgb}{0.8,0.8,0.8}
\definecolor{red}{rgb}{0.8,0.0,0.0}
\definecolor{LightCyan}{rgb}{0.88,1,1}

\begin{document}

\title{Black hole mergers of AGN origin in LIGO/Virgo's O1-O3a observing periods}
\author{V. Gayathri}
\affiliation{Department of Physics, University of Florida, PO Box 118440, Gainesville, FL 32611-8440, USA}
\author{Y. Yang}
\affiliation{Department of Physics, University of Florida, PO Box 118440, Gainesville, FL 32611-8440, USA}
\author{H. Tagawa}
\affiliation{Astronomical Institute, Graduate School of Science, Tohoku University, Aoba, Sendai 980-8578, Japan}
\author{Z. Haiman}
\affiliation{Department of Astronomy, Columbia University, 550 W. 120th St., New York, NY, 10027, USA}
\author{I. Bartos}
\thanks{imrebartos@ufl.edu}
\affiliation{Department of Physics, University of Florida, PO Box 118440, Gainesville, FL 32611-8440, USA}

\begin{abstract}
The origin of the black hole mergers detected by LIGO and Virgo remains an open question. While the unusual mass and spin of a few events constrain their possible astrophysical formation mechanisms, it is difficult to classify the bulk of the observed mergers. Here we consider the distribution of masses and spins in LIGO/Virgo's first and second observing catalogs, and find that for a significant fraction (25\%) of these detected events, an AGN-disk origin model is preferred over a parametric mass-spin model fit to the full GWTC-2 merger sample (Bayes factor $\mathcal{B}>10$). We use this to estimate the black hole merger rate in AGNs to be about $2.8\pm 1.8$\,Gpc$^{-3}$yr$^{-1}$, comparable to theoretical expectations. We find that AGNs can explain the rate and mass distribution of the observed events
with primary black hole mass in the pair-instability mass gap ($M\gtrsim50$\,M$_\odot$).
\end{abstract}

%%%%%%%%%%%%%%%%%%%%%%%%%%%%%%%%%%%%%%%%%
\section{Introduction}
%%%%%%%%%%%%%%%%%%%%%%%%%%%%%%%%%%%%%%%%%

Binary black holes can be the end results of several distinct astrophysical processes. They can form from isolated stellar binaries \citep{Bethe_1998,1998A&A...332..173P,2002ApJ...572..407B,2016A&A...588A..50M,2016MNRAS.460.3545D} or triples \citep{2014ApJ...781...45A,2016MNRAS.463.2443K,2020MNRAS.498L..46V},  dynamical interactions in star clusters \citep{1993Natur.364..423S,2000ApJ...528L..17P}, primordial black holes formed in the early universe \citep{10.1093/mnras/168.2.399}, and in the accretion disks of active galactic nuclei (AGNs; \citealt{2012MNRAS.425..460M,2017ApJ...835..165B,2017MNRAS.464..946S,2020ApJ...898...25T}). 

The origin of black hole mergers discovered by the LIGO \citep{2015CQGra..32g4001L} and Virgo \citep{2015CQGra..32b4001A} gravitational-wave observatories is not yet known \citep{2018arXiv181112907T,2020arXiv201014527A}. Nonetheless, some initial clues have emerged that challenge the isolated stellar binary origin of at least some of the events. These clues include the detection of at least one black hole in the pair-instability mass gap (50\,M$_\odot\lesssim M \lesssim 120$\,M$_\odot$; \citealt{2015ASSL..412..199W}) that may not be populated through isolated stellar evolution \citep{PhysRevLett.125.101102} (although supernova theory remains uncertain; \citealt{Belczynski:2020,DiCarlo:2020,Farmer:2020}); the detection of a compact object in the hypothesized lower mass gap ($2.5\,$M$_\odot\lesssim M \lesssim 5$\,M$_\odot$; \citealt{1998ApJ...499..367B,2012ApJ...757...55O}) between known neutron stars and black holes \citep{GW190814}; and the detection of mergers with large black hole spins that are misaligned from the binary orbital axis \citep{2020arXiv201014533T}.

To probe the origin of binary black hole mergers detected by LIGO/Virgo, one can compare the observed binaries' properties to expectations from the different astrophysical formation mechanisms. Such a comparison is, however, difficult due to the large number of possible scenarios, their unknown relative importance and their model uncertainties.  

We probed the contribution of the AGN formation channel to the detected population of black hole mergers. We carried out a Bayesian model comparison in which we contrasted the likelihood of each merger originating in an AGN disk to the likelihood of origin from the empirical mass-spin distribution obtained from the full observed black hole population. This strategy has the advantages that it only depends on one astrophysical model, and is conservative since the parameter distribution of the overall population can include AGN-assisted mergers as well.

\section{Model Comparison}

We carried out model comparisons for individual gravitational wave events. For merger population model $A$, we computed the probability distribution $P_{\rm pop}(\vec{\theta}|A)$ of binary parameters $\vec{\theta}$. Within $\vec{\theta}$ we considered the binary's chirp mass ${\cal{M}}=(m_1m_2)^{3/5}(m_1+m_2)^{-1/5}$, mass ratio $q=m_2/m_1$, effective spin $\chi_{\rm eff}= c G^{-1} (m_1+m_2)^{-1}(\vec{S}_1/m_1+\vec{S}_2/m_2)\vec{L}/|\vec{L}|$, 
which describes the objects' spin component parallel to the binary's orbital axis, and precessing spin $\chi_{\rm p}$ that describes the projection of component
spin vectors perpendicular to the orbital axis \citep{PhysRevLett.113.151101}. Here, $m_1$ and $m_2$ are the masses of the compact objects in the binary with $m_1>m_2$, $\vec{S}_1$ and $\vec{S}_1$ are the two objects' spin vectors, and $\vec{L}$ is the binary's orbital angular momentum vector.

We computed the posterior probability of $A$ given recorded gravitational wave data $\vec{x}$ for a single event as \citep{2019MNRAS.486.1086M,2020arXiv200705579V,2020ApJ...890L..20G}
\begin{equation}
    P(A|\vec{x}) =\pi(A) \frac{\int d \vec{\theta} P(\vec{\theta}|\vec{x})\pi(\vec{\theta})^{-1}P_{\rm pop}(\vec{\theta}|A)}{\int d\vec{\theta} P_{\rm det}(\vec{\theta})\, P_{\rm pop}(\vec{\theta}|A)}.
\end{equation}
where $\pi(A)$ is the prior probability of model $A$, $P(\vec{\theta}|\vec{x})$ is the probability density of true event parameters $\vec{\theta}$ for an observed gravitational wave event with observed data,
$\vec{x}$, $P_{\rm det}(\vec{\theta})$ is the detection probability for an event with true parameters $\vec{\theta}$,
$\pi(\vec{\theta})\propto d_{\rm L}(\vec{\theta})^2$ is the prior probability density of $\vec{\theta}$ at $d_{\rm L}$ luminosity distance \citep{2015PhRvL.115n1101V}.
 
We computed the Bayes factor for a given gravitational wave event for population models $A$ and $B$ for a give parameter space $\vec{\theta}$ as 
\begin{equation}
\mathcal{B}_{\rm B,A} = \frac{P(\vec{x}|B)}{P(\vec{x}|A)} = \frac{P(B|\vec{x})}{P(A|\vec{x})}\frac{\pi(B)}{\pi(A)}.
\end{equation}
A greater Bayes factor indicates that model $B$ is more likely to explain the observed data. 

%%%%%%%%%%%%%%%%%%%%%%%%%%%%%%%%%%%%%%%%%%%%%%%%%%%%%%%%%%%%
\section{Model 1: binary mergers in Active Galactic Nuclei}
%%%%%%%%%%%%%%%%%%%%%%%%%%%%%%%%%%%%%%%%%%%%%%%%%%%%%%%%%%%%

Following \cite{2017ApJ...835..165B}, we adopted a geometrically thin, optically thick, radiatively efficient, steady-state accretion disk expected in AGNs. We used a viscosity parameter $\alpha=0.1$ and a radiative efficiency $\epsilon=0.1$. We adopted a fiducial supermassive black hole mass $M_{\bullet}=10^6$\,M$_\odot$ and accretion rate $0.1\dot{M}_{\rm Edd}$, where $\dot{M}_{\rm Edd}$ is the Eddington accretion rate. 

We computed the expected mass and spin distributions of binary mergers in AGNs following \cite{2020ApJ...901L..34Y} and \cite{2020ApJ...898...25T,2020ApJ...899...26T}. For simplicity, we adopted a Salpeter initial mass function $dN/dm\propto m^{-2.35}$ for black holes and a normal initial mass function $m/M_{\odot}\sim\textit{N}(1.49,0.19)$ for neutron stars \citep{2016ARA&A..54..401O}. We assumed that the total mass of the black hole population is $1.6\%$ of the stellar mass in galactic centers and the number of neutron stars is ten times the number of stellar-mass black holes. We also took into account the mass segregation in the spatial distributions of black holes and neutron stars following O'Leary et al. \citep{2009MNRAS.395.2127O, 2018ApJ...860....5G}.

Black holes and neutron stars orbiting the central supermassive black hole  will  periodically cross the AGN disk. We simulated the process of orbital alignment with the disk following the method of \cite{2019ApJ...876..122Y} by simulating the orbital evolution of $10^5$ black holes and neutron stars.

Neutron stars and stellar-mass black holes were assumed to migrate from their original locations inward once they have been aligned with the AGN disk \citep{2020ApJ...899...26T}. The type \uppercase\expandafter{\romannumeral1} and type \uppercase\expandafter{\romannumeral2} time scale for migration was taken to be \citep{2010MNRAS.401.1950P,2011ApJ...726...28B,2002ApJ...565.1257T,2014ApJ...792L..10D,2014ApJ...782...88F,2015ApJ...806L..15K,2018ApJ...861..140K}
\begin{align}
    t_{\rm \uppercase\expandafter{\romannumeral1}}=&\frac{1}{2f_{\rm mig}}\frac{M_{\bullet}}{M_{\rm bh}}\frac{M_{\bullet}}{\Sigma r^2}\left( \frac{H}{r}\right)^2\Omega^{-1}\\
    t_{\rm \uppercase\expandafter{\romannumeral1}/\uppercase\expandafter{\romannumeral2}}=&(1+0.04K)t_{\rm \uppercase\expandafter{\romannumeral1}}
\end{align}
Here, $f_{\rm mig}=2$ is a dimensionless factor and $\Omega$ is the Keplerian angular velocity of an orbit with radius $r$, $K=(M_{\rm bh}/M_{\bullet})^2(H/r)^{-5}\alpha^{-1}$ and $\alpha$ is the viscosity parameter. The equation for migration of the objects in the AGN disk is $dr/dt=-r/t_{\rm I/II}$ \citep{2020ApJ...898...25T}.

Neutron  stars  and  stellar-mass  black holes can undergo close encounters with other objects around the supermassive black hole. When this happens in the AGN disk, the surrounding gas might remove enough energy such that the two objects involved in the close encounter are able to form a binary \citep{Goldreich2002,2020ApJ...898...25T}. %HT5: I added a reference of Goldreich+02
We found through numerical simulations that the flux of black holes or neutron stars passing by an object in the AGN disk can be approximated by:
\begin{equation}
    \phi_{i}=2\times10^{-8} \,\mbox{AU}^{-2}{\rm yr}^{-1} \left(\frac{\rho_{i}}{5000 \,\mbox{pc}^{-3}}\right)
\end{equation}
where the index $i$ can refer to black holes or neutron stars, and $\rho_{i}$ is the number density of black holes or neutron stars in the galactic center. The orientation of their velocity was assumed to be isotropically distributed, following a normal distribution $\textit{N}(0,\sigma^2_{\rm v})$ in each spatial dimension, where $\sigma^2_{\rm v}=2.5\times10^{3}(r/{\rm pc})^{-1}(\rm km/s)^2$. 
The rate for close encounters in the AGN disk is then $\Gamma_{\rm en,\rm i}\simeq\sqrt{3}\phi_{i}r^2_{\rm H}\langle v_{\rm rel} \rangle/ \sigma_{\rm v}\sim \sqrt{3}\phi_{i}r^2_{\rm H}\sqrt{1+v^2_{\rm kep}/\sigma^2_{\rm v}}$, where $r_{\rm H}$ is the mutual Hill radius and $v_{\rm kep}$ is the Keplerian velocity of  an  orbit  with  radius $r$. When the two objects have small relative velocity, the gaseous friction is able to remove the necessary energy for binary formation \citep{2020ApJ...898...25T}. However, in our case, we assumed that the initial orbital orientations of the unbound compact objects are random. This makes the relative velocity of the two objects comparable to the Keplerian velocity, which can greatly reduce the probability of the two objects getting bounded. The probability to form a binary during each close encounter can be approximated by \citep{2020ApJ...898...25T}
\begin{equation}
P_{\rm cap}=t_{\rm pass}/t_{\rm GDF}\simeq 4\pi G^2 m_{\rm t}\rho_{\rm gas}r_{\rm H}v^{-4}_{\rm rel}f\left(\frac{v_{\rm rel}}{c_{\rm s}}\right)\nonumber,
\end{equation}
where $f(v_{\rm rel}/c_{\rm s})$ accounts for deceleration due to dynamical friction, $ m_{\rm t}$ is the total mass, $\rho_{\rm gas}$ is the gas density, $r_{\rm H}$ is the Hill radius and $t_{\rm GDF}$ is the gas dynamical friction hardening
timescale. 

Binaries can also form dynamically via  three-body interaction \citep{2008gady.book.....B}. We assumed that a binary can form when three objects undergo a strong encounter and adopted a rough estimate of the formation rate
\begin{equation}
    \Gamma_{\rm 3bbf,\rm i}\simeq\frac{1}{2}n^2_{\rm i}b^5_{\rm str}v_{\rm rel}\simeq\frac{1}{2}{\phi}^2_{\rm i}b^5_{\rm str}/v_{\rm rel}
\end{equation}
where $n_{\rm obj}$ is the density of black holes or neutron stars in the galactic center and $b_{\rm str}=\min\{b_{90}, r_{\rm H}\}$, $b_{90}=Gm_{\rm tot}/v^2_{\rm rel}$. Since the relative velocity $v_{\rm rel}$ is on the order of the Keplerian velocity, $b_{90}$ is much smaller than the Hill radius $r_{\rm H}$ and thus the rate of formation via three-body encounters $\Gamma_{\rm 3bbf}$ is negligible compared with the gas-capture binary formation rate $\Gamma_{\rm cap}\equiv P_{\rm cap}\Gamma_{\rm en}$. With the above assumptions, gas-capture typically occurs in the inner regions ($\sim10^{-4}-10^{-2}$\,pc) of the AGN disk where gas density is high.

%\subsubsection{Angular momentum evolution}
Stellar-mass black holes capture gas from the AGN disk while they are moving in or crossing the disk. When the BHs accrete this gas, their masses and spins evolve. The spin magnitude after $\Delta m_{\rm BH}$ being accreted onto the BH is given by \citep{1970Natur.226...64B,2020ApJ...899...26T}
\begin{equation}
    a^{\rm f}=\frac{1}{3}\frac{r^2_{\rm isco}}{f_{\rm acc}}[4-(3\frac{r_{\rm isco}}{f^2_{\rm acc}}-1)^{1/2}]
\end{equation}
where $f_{\rm acc}=(m_{\rm BH}+\Delta m_{\rm BH})/m_{\rm BH}$ and $r_{\rm isco}$ is the  radius of the innermost stable circular orbit (ISCO) in gravitational units. The BHs possibly possess mini-disks, if the spin angular momentum $\vec{J}_{\rm BH}=\vec{a}\sqrt{G m^3_{\rm BH} R_{\rm g}}$ of a BH is misaligned with its inner disk, they will tend to align due to Lense-Thirring precession. The inner parts of the mini-disk is defined by a so-called  warp radius $R_{\rm warp}$ which is given by \citep{Volonteri_2007}
\begin{equation}
    R_{\rm warp}/R_{\rm g}=3.6\times10^2|\vec{a}|^{5/8}m^{1/8}_{\rm BH}\lambda^{-1/4}(\frac{\nu_2}{\nu_1})\alpha^{-1/2}
\end{equation}
here, $\lambda$ is the Eddington ratio, $\nu_1$
is the viscosity corresponding to angular momentum transfer in the accretion disk and $\nu_2$ is the viscosity responsible for warp propagation. We assumed that the spin of a BH can align with the total angular momentum $\vec{J}_{\rm tot}=\vec{J}_{\rm BH}+\vec{J}_{\rm warp}$ in each time step of our simulations, where $J_{\rm warp}\simeq \Delta m_{\rm BH} \sqrt{G m_{\rm BH} R_{\rm warp}}$. Following \citet{2020ApJ...899...26T}, we assumed that the direction of $\vec{J}_{\rm warp}$ is aligned with AGN disk angular momentum for a single BH and is aligned with the orbital angular momentum for a BH in binary. In our fiducial model, we assumed $\nu_2/\nu_1=10$.

The direction of orbital angular momentum of binaries evolve due to accretion torques. We assumed that the orbital angular momentum after $\Delta m_{\rm Bin}$ being captured by the binary is $\vec{J^{\rm f}}_{\rm bin}=\vec{J}_{\rm bin}+\vec{J}_{\rm gas}$ \citep{2020ApJ...899...26T,1999ApJ...526.1001L}
\begin{equation}
\vec{J}_{\rm gas}\simeq \Delta m_{\rm Bin}\sqrt{Gm_{\rm bin}s}\hat{J}_{\rm AGN}    
\end{equation}
where $s$ is the separation of the binary
and $\hat{J}_{\rm AGN}$ is the direction of AGN disk angular momentum.

%\subsubsection{Binary separation evolution}
When a binary travels in the AGN disk, the surrounding gas can provide a drag force on the binary, which will reduce its separation. We followed the results of \cite{2007ApJ...665..432K,Kim_2008} for dynamical force in gaseous medium and assumed that the drag force is $F_{df}\mathcal{M}^2=const.$ when $\mathcal{M}>8$, where $\mathcal{M}$ is the Mach number. More recent simulations of binary evolution give similar results within a factor of 2 \citep{2019ApJ...884...22A,2021arXiv210312088K}.

When the orbital separation of a binary is sufficiently compact, gravitational radiation dominates the dynamical drag force. The hardening rate of the binary due to a GW radiation is
\begin{equation}
    \frac{ds}{dt}=-\frac{64}{5}\frac{G^3m_1m_2(m_1+m_2)}{c^5s^3}. 
\end{equation}
We assumed that the actual hardening rate is the combination of the contribution from dynamical friction and GW radiation. We did not take into account stellar interactions such as binary-single interactions, which are important if a gap is opened and result in the accumulation of many binaries and compact objects \citep{2020arXiv201200011T}. In our fiducial model adopted here, most binaries and compact objects will not open a gap, making binary-single interactions less important. 

The spatial distribution in galactic nuclei is not yet well constrained. While we assumed above an isotropic initial distribution, vector resonant relaxation could substantially reduce the black holes' initial velocity dispersion \citep{Szolgyen2018}, 
which can increase the chance of capture during a close encounter as well as the probability of forming binaries via three-body encounters. To characterize this scenario, we additionally considered below the AGN-assisted merger model of \cite{2020arXiv201200011T} that adopts a small initial velocity dispersion of $\sim 0.2v_{\rm kep}$.  Beyond the difference in the velocity dispersion, \cite{2020arXiv201200011T} also adopts a black hole initial mass function with maximum mass of $15\,$M$_\odot$ (c.f. $50$\,M$_\odot$ for the fiducial model), reflecting the high metallicity observed near AGN disks.

\begin{figure*}
%HT5: I think "Tagawa-20" in the legend should be revised to "Tagawa-21" as the distributions are produced by the model updated in Tagawa+21. 
\includegraphics[scale=0.42]{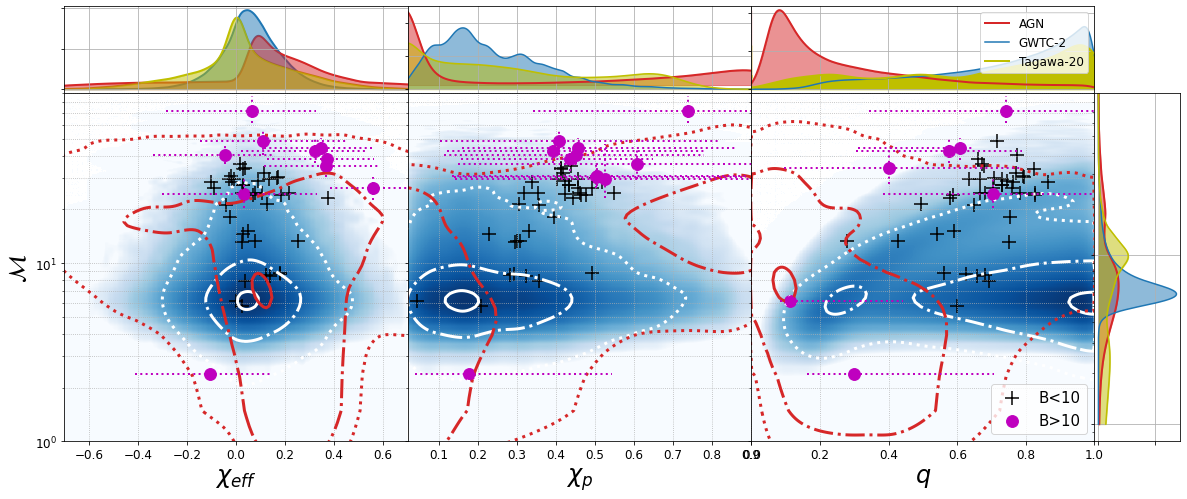}%mc_chi_eff_chi_p_q_with_BF_agn_gwtc2_tagawa_err_contour_10_march15.png}
\caption{{\bf Probability densities for the GWTC-2 model and reconstructed parameters for gravitational wave events}. The distributions and reconstructed values are shown for ${\cal{M}}-\chi_{\rm eff}$ (left), ${\cal{M}}-\chi_{\rm p}$ (middle) and ${\cal{M}}-q$ (right). Probability densities for single parameters marginalized over other parameters are also shown on the top and on the right 
%ZH2: is the distribution shown on the right marginalized over x_eff and x_p?   or  also over q?   please define.
% gayathri: Yes, it marginalized over other parameters. Incorporated in caption. 
%ZH3: thanks -- i was specifically asking about the right panel, where the M_chirp distribution is shown.
%     is this marginalized over "q" only?   i guess maybe that's implied because it's on the right panel 
%     (but would be good to state unambiguously).
%     Also these are all 2D analyses, right?  Or are they in 3D?  I.e. the 1-D distributions have been marginalized 
%     over 1 or 2 additional parameters?   For example, the top of the left panel shows the M-distribution.  Was this
%     marginalized over X_eff only (i.e. it's still within the 2D slice)?  Or was it marginalized over both X_eff and X_p?
%Gayathri: all one 1d histograms are marginalised over other parameters like chirp_mass hist marginalised over q, chi_p and chi_eff. 
% Then 2-D histograms is marginalised over other two parameters, like chirp_mass - chi_eff contour marginalised over q and chi_p. 
%     
for the GWTC-2 model, for the fiducial AGN model in this work, and for the AGN model of \cite{2020arXiv201200011T} (see legend). Gravitational wave events with Bayes factor $\mathcal{B}>10$ ($\mathcal{B}<10$) are shown in magenta (black). For the $\mathcal{B}>10$ case we also show the 90\% credible intervals of the reconstructed parameters (dotted lines). For the GWTC-2 model probability density, white lines mark the 95\%,  65\% and 10\% credible areas. The red lines mark the 95\%,  65\% and 10\% credible areas for the fiducial AGN model in this work. Note that the shown probability densities depict the expected rate of occurrence, i.e. they are not weighted by the expected detection volume, while the distributions of the parameters of observed events are affected by their detection volumes.}
\label{fig:pdf_with_gw_events}
\end{figure*}

%%%%%%%%%%%%%%%%%%%%%%%%%%%%%%%%%%%%%%%%%%%%%%%%%%%%%%%%%%
\section{Model 2: LIGO/Virgo's GWTC-2 distribution} \label{se:GWTC-2}
%%%%%%%%%%%%%%%%%%%%%%%%%%%%%%%%%%%%%%%%%%%%%%%%%%%%%%%%%%

To differentiate AGN-assisted mergers from the rest of LIGO/Virgo's detections, we considered the population properties of the 47 binary mergers identified by LIGO/Virgo for the available observing runs O1, O2 and O3a \citep{2018arXiv181112907T,2020arXiv201014527A}. We adopted the obtained parameter distributions for the "Power Law + Peak mass" model of \cite{2020arXiv201014533T} (hereafter the GWTC-2 model). For $m_1$ this model gives 
\begin{eqnarray}\label{eq-primarymass}
\pi(m_1 | \lambda_\text{peak}, \alpha, m_{min}, \delta_m, {m_{max}}, \mu_m, \sigma_m) =  \quad\quad\quad\quad \nonumber \\
(1-\lambda_\text{peak})\mathfrak{P}(m_1|-\alpha, m_{max})S(m_1|m_{min}, \delta_m) \nonumber \\ + \lambda_\text{peak} G(m_1|\mu_m,\sigma_m) 
 S(m_1|m_{min}, \delta_m),  \quad \quad
\end{eqnarray}
Here, the three functions on the right-hand side describe a normalized power-law distribution with spectral index $-\alpha$ and high-mass cut-off $m_{max}$ ($\mathfrak{P}$),  a normalized Gaussian distribution with mean $\mu_m$ and width $\sigma_m$ ($G$), and a smoothing function which rises from 0 to 1 between $m_{min}$, and $m_{min}+\delta_m$ ($S$).  The parameter $\lambda_\text{peak}$ is a mixing fraction determining the relative prevalence of mergers in power-law and Gaussian distribution.

We adopted the conditional mass ratio distribution for this model,
\begin{align}
\label{eq:pq_smoothing}
\pi(q \mid \beta, m_1, m_{min}, \delta_m) \propto q^{\beta_q} S(q m_1 \mid m_{min}, \delta_m).
\end{align}

The mass distribution model is represented by eight parameters namely $\lambda_{peak}$, $\alpha$, $m_{min}$, $\delta_m$, $m_{max}$, $\mu_m$, $\sigma_m$, and $\beta_q$. Here, $m_1$ and $m_2$ distributions were generated from hyper-parameter space these parameters by reported by \cite{2020arXiv201014533T} (see their Fig. 16). 

Similarly, we considered  their default spin model for spin estimation. This dimensionless spin magnitude model is represented by two parameters namely $\alpha_{\chi}$ and $\beta_{\chi}$ by a Beta distribution, $\pi(\chi_{1,2}|\alpha_{\chi},\beta_{\chi})={\rm Beta}(\alpha_x,\beta_x)$. These parameters are known as standard shape parameters that determine the distribution’s mean and variance. Here we assume same distribution for $\chi_1$ and $\chi_2$. The cosine of the tilt angle between component spin and binary orbital angular momentum is represented by two parameters namely $\zeta$ and $\sigma_{t}$. This cosine angle is distributed as a mixture of two populations ( $\cal{J}$ isotropic distribution +  $G_t$ truncated Gaussian), $\pi(\cos \, \theta_{1,2}|\zeta,\sigma_{t})=\zeta G_t(\cos \, \theta_{1,2}|\sigma_{t})+ (1-\zeta){\cal{J}}(\cos \, \theta_{1,2})$.

%%%%%%%%%%%%%%%%%%%%%%%%%%%%%%%%%%%%%%%%%%%%
\section{Results}
%%%%%%%%%%%%%%%%%%%%%%%%%%%%%%%%%%%%%%%%%%%%

We computed the Bayes factor for each binary merger detected by LIGO/Virgo in their O1, O2 and O3a observing runs by comparing our AGN model to the baseline GWTC-2 model.

We determined the Bayes factor using the triplet of parameters $\cal{M}$, $\chi_{\rm eff}$, and $\chi_{\rm p}$. We omitted the mass ratio $q$ 
due to the computational cost of the 4D analysis. As a result, we have considered merger population distribution marginalized over $q$.  We also determined the Bayes factor using pairs of parameters involving the chirp mass $\cal{M}$ and either of the effective spin $\chi_{\rm eff}$, precessing spin $\chi_{\rm p}$, or mass ratio $q$.  Cases in which the parameter $q$ favors the AGN model can be gauged by ${\cal{M}}-q$ Bayes factor. 

%HT5: I note that the number for the first table becomes 6 instead of 1. 
Our results are shown in Table \ref{tab:1}. We see that 12 out of the 47 gravitational wave events have Bayes factor $\mathcal{B}>10$, which we use here as a threshold to indicate probable AGN origin. 

To show the parameters of the binaries that are distinct from the GWTC-2 model and are likely of AGN origin, in Fig. \ref{fig:pdf_with_gw_events} we show 2D slices of the parameter distributions. These slices include the  ${\cal{M}}-\chi_{\rm eff}$, ${\cal{M}}-\chi_{\rm p}$ and ${\cal{M}}-q$ combinations. For each combination we show the GWTC-2 distribution and the reconstructed parameters for LIGO/Virgo's O1, O2 and O3a events, highlighting the ones with Bayes factor $>10$. The numerical values of the Bayes factors for the above three combinations are listed in Table \ref{tab:1}. 

We see that the AGN model is favored mainly for high-mass, high-spin sources, where the GWTC-2 model has weaker support. Note that the AGN model of \cite{2020arXiv201200011T} generally favors lower masses as its BH initial mass function extends only up to $15$\,M$_\odot$, and accordingly high-mass binaries have less support for this model. 
In the AGN model, $\chi_\mathrm{eff}$ is distributed in low values due to the formation of binaries between black holes in and outside of the AGN disk, which reproduces random directions for the angular momentum directions of binaries, while low $\chi_\mathrm{eff}$ is reproduced by frequent binary-single interactions in the model of \cite{2020arXiv201200011T}.

With the above results we estimated the overall fraction of AGN-assisted mergers within LIGO/Virgo's detected sources as follows. We considered a chirp mass threshold of ${\cal{M}}_{\rm th} = 40$\,M$_\odot$ above which most (6 out of 7) events are favored to have an AGN origin with Bayes factor $\mathcal{B}>10$. We then computed the overall AGN-assisted merger rate that corresponds to 6 expected detections above this mass from AGNs by LIGO/Virgo during O1, O2 and O3a. In this computation we adjusted the overall AGN merger rate density in comoving volume. We assumed a uniform rate density distribution, reflecting the shallow cosmic evolution of the merger rate \citep{Yang_2020}. All other model parameters were adopted as described above in our fiducial model, and were not adjusted in this fit. The expected number of detections was calculated using LIGO/Virgo's noise curve for the three observing runs, the observing times, and the binary gravitational waveforms as functions of binary mass and spin. We required a network-wide signal-to-noise ratio of 8 for detection. With this we found the AGN BH merger rate to be about $2.7\pm1.8$\,Gpc$^{-3}$yr$^{-1}$, where the error bars correspond to the 90\% confidence interval accounting for statistical uncertainty. This corresponds to an overall expected $15\pm10$ (90\% confidence interval) binary mergers of AGN origin in LIGO/Virgo's O1-O3a detected sample, which is comparable to the number of events we identified with Bayes factor $\mathcal{B}>10$.
%HT4: I see. But is the AGN density distribution specified?  IB: if you mean the number density of AGNs this is folded in in our estimate of uniform rate density, it doesn't appear independently in the calculations.
%HT5: I'm still not sure about this. Did you adopt the AGN luminosity function from Shen+2020 as adopted in Yang+2020, or use that from Greene and Ho 2007 as adopted in Bartos+2017, or adopt 0.018Mpc^-3 from Hao+2005 as adopted in Yang+2019, or assume some specific value, which is adequate for a 10^6Msun SMBH with the 0.1 Eddington accretion rate? 

\begin{figure*}
\begin{center}
\hspace{-0.2cm}
%HT5: As in Fig. 1, "Tagawa-20" in the legend is good to be revised to "Tagawa-21" (I already revised these in captions). 
\includegraphics[scale=0.48]{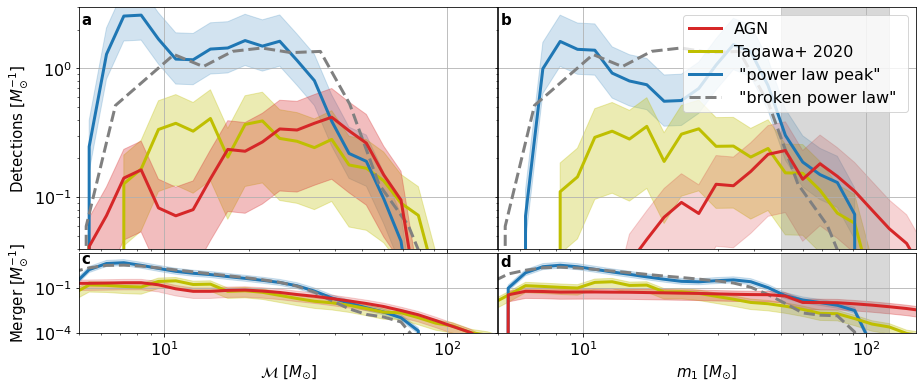}
\caption{{\bf Detection and merger spectra of $\mathcal{M}$ and $m_1$ for the AGN and GWTC-2 models.} Expected number of detections (a,b) and expected merger rate (c,d) per unit mass as functions of $\mathcal{M}$ (a,c) and $m_1$ (b,d) for the AGN model (red), Tagawa model+2021 (yellow), and the GWTC-2 model (blue). 
%HT5: I guess error regions just reflects the errors in the estimated rate of 2.7+-1.8Gpc^-3yr^-1, right? I suppose it is good to be mentioned. 
For our fiducial GWTC-2 model we adopted the ``power law peak" model from \cite{2020arXiv201014533T}. For comparison, we also show the ``broken power law" model from \cite{2020arXiv201014533T} (gray dashed, see legend). In panels (a) and (b) the vertical gray band marks the expected range of the black hole pair-instability mass gap $\sim50-120$\,M$_\odot$. 
%HT3: If it is not difficult, I'm really curious to see the distribution for a model in Tagawa+. (Sorry for the late ask)
%HT: Thanks a lot for adding it! It looks interesting and useful. 
}
\label{fig:rate}
\end{center}
\end{figure*}
%\end{widetext}

In Fig. \ref{fig:rate} we show the expected number of detections per unit mass as a function of $\mathcal{M}$ and $m_1$ both for the AGN and for GWTC-2 models. For the AGN model, this spectrum was normalized so it corresponds to a total of 6 detections for $\mathcal{M}>40$\,M$_\odot$ and a corresponding 15 detections for all masses. For the GWTC-2 model the curve corresponds to a total of 47 detections, i.e. the actual number of LIGO/Virgo's detections during O1-O3a. 
For comparison, we also show the corresponding detection and merger spectrum for an alternative GWTC-2 model that fits a broken power law on the detected masses \citep{2020arXiv201014533T}, and
we also mark the pair-instability mass gap of $50-120$\,M$_\odot$ \citep{2017ApJ...836..244W} by a gray band.

We see that, below 50\,M$_\odot$, the GWTC-2 spectrum is substantially above the spectrum of AGN detections. This relation changes above $\sim50$\,M$_\odot$, where the AGN channel dominates, especially above the maximum mass $\sim90$\,M$_\odot$ allowed by the GWTC-2 model.

\section{Conclusions}

We carried out a Bayesian model comparison to probe which of LIGO/Virgo's binary mergers detected within the O1, O2 and O3a observing periods are most likely to be of AGN origin. We used the one fiducial model from GWTC-2 as a conservative comparison as it is a fit to the observed data and therefore also includes events that are possibly of AGN origin. We used the obtained AGN vs GWTC-2 Bayes factors to examine the population of AGN-assisted mergers within the population detected by LIGO/Virgo. Our conclusions are summarized below:
\begin{itemize}[leftmargin=*,itemsep=1pt,topsep=1pt]
\item Out of the 47 events in the GWTC-2 catalog, 12 have Bayes factor $\mathcal{B}>10$, i.e. are better fit by the AGN disk model than by the GWTC-2 distribution.
\item Using the highest-mass events, which are all better explained by the AGN model, we estimate the total black hole merger rate in AGNs to be $2.7\pm1.8$\,Gpc$^{-3}$yr$^{-1}$ ($90\%$ confidence level statistical uncertainty). This is $2-30$\% of the total merger rate $23.9^{+14.9}_{-8.6}$\,Gpc$^{-3}$yr$^{-1}$ estimated from all LIGO/Virgo detections. 
\item The detected mass distribution expected from the AGN channel reproduces well (without any fit parameters) the GWTC-2 distribution fit to observations at high masses (${\cal{M}}\gtrsim 40$\,M$_\odot$ or $m_1\gtrsim 50$\,M$_\odot$). AGNs only marginally contribute to the detected lower-mass mergers ($5$\,M$_\odot\lesssim m\lesssim 50$\,M$_\odot$), therefore, the mass spectrum of black holes in this mass range is representative of non-AGN formation channels observed by LIGO/Virgo.
\end{itemize}
We note several caveats regarding the above conclusions. 
First, the fitting function used to model the detected LIGO/Virgo sample for the GWTC-2 model \citep{2020arXiv201014533T} may have too few parameters to accurately fit all properties of the true mass and spin distributions, which could result in a too high Bayes factor favoring the AGN channel for events with high mass and/or spin. 
Second, other merger channels, for example hierarchical mergers in non-AGN environments, could similarly result in a black hole mass distribution that extends to high masses and accounts for some of the most massive events we associate here with AGNs. 
Third, our results depend on the AGN model parameters we adopted for or simulation. Adopting the model of \citet{2020arXiv201200011T}, which produces mergers with low binary masses, we found no source with high-Bayes factor preference for an AGN origin
except GW190426\_152155. 
Fourth, here we did not take into account the fact that some AGN-assisted mergers could be highly eccentric \citep{2020arXiv201009765S,2021ApJ...907L..20T}, which would affect their detectability and the accuracy of their mass and spin reconstruction. These caveats merit further study. Similarly, other approaches also used to analyze the hierarchical component are discussed in \cite{2020ApJ...893...35D,McKernan:2019beu,2020arXiv201014533T}.

Nonetheless, assuming that the above described AGN contribution is accurate, we can make some predictions on the expected features of mergers and future detections:
\begin{itemize}[leftmargin=*,itemsep=1pt,topsep=1pt]
\item Future observations should uncover mergers with black hole masses $>100$\,M$_\odot$, which could represent up to a few percent of the detected events. Many of these exceptionally massive events should have high precessing spins. 
\item The large fractional contribution of AGNs enhances the utility of correlating the localization of the detected mergers with catalogs of AGNs. A 30\% contribution may be identifiable from several hundred detections (all channels included), assuming sufficiently complete AGN catalogs  \citep{2017NatCo...8..831B}.
\end{itemize}

\renewcommand{\arraystretch}{0.95}
\begin{table*}
\begin{center}
\label{tab:1}
\begin{tabular}{|p{2.5cm}|p{1.4cm}|p{1.3cm}|p{1.3cm}|p{1.3cm}|p{1.3cm}|p{1.2cm}|p{0.7cm}|p{0.6cm}|p{0.6cm}|p{0.6cm}|p{0.6cm}|}
\hline
\multirow{2}{*}{\bf{Event}} &\multicolumn{6}{c|}{Estimated source parameters} & \multicolumn{5}{c|}{Bayes factor} \\ \cline{2-12}
 
& ${m_1}$ & ${m_2}$& $\cal{M}$ & $q$& $\chi_{\rm eff}$& $\chi_{\rm p}$ &  $\cal{M}$$\chi_{\rm eff}$ & $\cal{M}$$\chi_{p}$ &  $\cal{M}$$q$  & \multicolumn{2}{c|}{$\cal{M}$$\chi_{eff}$$\chi_{p}$}  \\

 \hline %1
GW150914 &$ 35.55 ^{+ 4.66 } _{ -3.1 }$
&$ 30.52 ^{+ 3.02 } _{ -4.38 }$
&$ 28.52 ^{+ 1.67 } _{ -1.46 }$
&$ 0.86 ^{+ 0.12 } _{ -0.2 }$
&$ -0.01 ^{+ 0.12 } _{ -0.13 }$
&$ 0.34 ^{+ 0.45 } _{ -0.25 }$
&$ 1.5 $ &\bf{3.4}&$ 0.6 $ &$ 0.5 $ &$ 0.5 $ \\
\hline%2
GW170104 &$ 30.7 ^{+ 7.32 } _{ -5.58 }$
&$ 19.94 ^{+ 4.87 } _{ -4.56 }$
&$ 21.31 ^{+ 2.21 } _{ -1.78 }$
&$ 0.65 ^{+ 0.3 } _{ -0.23 }$
&$ -0.04 ^{+ 0.17 } _{ -0.21 }$
&$ 0.36 ^{+ 0.42 } _{ -0.27 }$
&\bf{1.9}&$ 2.3 $ &$ 0.9 $ &$ 0.9 $ &$ 2.2 $ \\
\hline%3
GW170809 &$ 34.81 ^{+ 8.25 } _{ -5.85 }$
&$ 23.73 ^{+ 5.12 } _{ -5.15 }$
&$ 24.79 ^{+ 2.14 } _{ -1.66 }$
&$ 0.68 ^{+ 0.28 } _{ -0.24 }$
&$ 0.08 ^{+ 0.17 } _{ -0.17 }$
&$ 0.35 ^{+ 0.43 } _{ -0.26 }$
&$ 1.2 $ &\bf{2.5}&$ 1.1 $ &$ 0.6 $ &$ 0.4 $ \\
\hline%4
GW170823 &$ 39.27 ^{+ 11.18 } _{ -6.68 }$
&$ 28.84 ^{+ 6.72 } _{ -7.79 }$
&$ 28.97 ^{+ 4.61 } _{ -3.6 }$
&$ 0.74 ^{+ 0.23 } _{ -0.3 }$
&$ 0.09 ^{+ 0.22 } _{ -0.26 }$
&$ 0.42 ^{+ 0.41 } _{ -0.31 }$
&$ 3.1 $ &\bf{5.6}&$ 1.9 $ &$ 3.5 $ &$ 1.1 $ \\
\hline%5
GW151012 &$ 23.1 ^{+ 14.9 } _{ -5.45 }$
&$ 13.56 ^{+ 4.06 } _{ -4.76 }$
&$ 15.15 ^{+ 2.08 } _{ -1.2 }$
&$ 0.59 ^{+ 0.36 } _{ -0.35 }$
&$ 0.05 ^{+ 0.31 } _{ -0.2 }$
&$ 0.33 ^{+ 0.45 } _{ -0.25 }$
&$ 0.7 $ &\bf{1.4}&$ 0.7 $ &$ 0.6 $ &$ 1.4 $ \\
\hline%6
GW170608 &$ 10.95 ^{+ 5.44 } _{ -1.71 }$
&$ 7.59 ^{+ 1.36 } _{ -2.18 }$
&$ 7.92 ^{+ 0.19 } _{ -0.18 }$
&$ 0.69 ^{+ 0.28 } _{ -0.36 }$
&$ 0.03 ^{+ 0.19 } _{ -0.07 }$
&$ 0.36 ^{+ 0.45 } _{ -0.27 }$
&\bf{0.7}&$ 0.3 $ &$ 0.2 $ &$ 0.3 $ &$ 0.5 $ \\
\hline%7
GW170814 &$ 30.5 ^{+ 5.57 } _{ -2.95 }$
&$ 25.14 ^{+ 2.81 } _{ -4.03 }$
&$ 24.02 ^{+ 1.4 } _{ -1.13 }$
&$ 0.83 ^{+ 0.15 } _{ -0.23 }$
&$ 0.07 ^{+ 0.12 } _{ -0.12 }$
&$ 0.48 ^{+ 0.41 } _{ -0.36 }$
&$ 0.8 $ &\bf{3.6}&$ 0.5 $ &$ 0.7 $ &$ 0.4 $ \\
\hline%8
GW151226 &$ 13.68 ^{+ 8.75 } _{ -3.23 }$
&$ 7.66 ^{+ 2.18 } _{ -2.54 }$
&$ 8.83 ^{+ 0.34 } _{ -0.29 }$
&$ 0.56 ^{+ 0.38 } _{ -0.33 }$
&$ 0.18 ^{+ 0.2 } _{ -0.12 }$
&$ 0.49 ^{+ 0.39 } _{ -0.32 }$
&\bf{1.0}&$ 0.5 $ &$ 0.2 $ &$ 0.3 $ &$ 0.5 $ \\
\hline%9
%\rowcolor{GRAY}
%\rowcolor{gray!50}
%\rowcolor{gray}
\textcolor{red}{{GW170729}} &\textcolor{red}{$ {49.84 ^{+ 16.17 } _{ -10.17 }}$}
& \textcolor{red}{\bf{$ 33.71 ^{+ 9.02 } _{ -9.98 }$}}
&\textcolor{red}{\bf{$ 35.15 ^{+ 6.48 } _{ -4.75 }$}}
&\textcolor{red}{\bf{$ 0.68 ^{+ 0.28 } _{ -0.28 }$}}
&\textcolor{red}{\bf{$ 0.37 ^{+ 0.21 } _{ -0.25 }$}}
&\textcolor{red}{\bf{$ 0.44 ^{+ 0.35 } _{ -0.28 }$}}
&\textcolor{red}{{\bf 15}}
&\textcolor{red}{$ 9.3 $} 
&\textcolor{red}{$ 5.4 $} 
&\textcolor{red}{$ 14 $} 
&\textcolor{red}{$ 1.3 $} \\
\hline%10
GW170818 &$ 35.2 ^{+ 7.43 } _{ -4.72 }$
&$ 26.6 ^{+ 4.28 } _{ -5.2 }$
&$ 26.42 ^{+ 2.12 } _{ -1.71 }$
&$ 0.76 ^{+ 0.21 } _{ -0.25 }$
&$ -0.09 ^{+ 0.18 } _{ -0.21 }$
&$ 0.49 ^{+ 0.37 } _{ -0.34 }$
&$ 4.2 $ &\bf{7.3}&$ 0.8 $ &$ 5.5 $ &$ 2.4 $ \\
\hline%11
%\hline
GW190408\_181802
&$ 24.46 ^{+ 5.04 } _{ -3.36 }$
&$ 18.32 ^{+ 3.16 } _{ -3.5 }$
&$ 18.25 ^{+ 1.75 } _{ -1.16 }$
&$ 0.76 ^{+ 0.21 } _{ -0.24 }$
&$ -0.03 ^{+ 0.13 } _{ -0.19 }$
&$ 0.39 ^{+ 0.38 } _{ -0.31 }$
&$ 0.8 $ &\bf{2.3}&$ 0.5 $ &$ 0.6 $ &$ 1.4 $ \\
\hline%48
GW190412 &$ 30.22 ^{+ 4.06 } _{ -4.1 }$
&$ 8.45 ^{+ 0.99 } _{ -0.92 }$
&$ 13.36 ^{+ 0.53 } _{ -0.53 }$
&$ 0.28 ^{+ 0.08 } _{ -0.06 }$
&$ 0.25 ^{+ 0.08 } _{ -0.08 }$
&$ 0.3 ^{+ 0.18 } _{ -0.15 }$
&$ 1.7 $ &$ 0.4 $ &\bf{4.9}&$ 0.5 $ &$ 2.0 $ \\
\hline%12
%\rowcolor{GRAY}
\textcolor{red}{GW190413\_052954}
& \textcolor{red}{$ 33.89 ^{+ 12.57 } _{ -7.56 }$}
&\textcolor{red}{$ 23.55 ^{+ 6.68 } _{ -6.48 }$}
&\textcolor{red}{$ 24.33 ^{+ 5.2 } _{ -3.83 }$}
&\textcolor{red}{$ 0.7 ^{+ 0.26 } _{ -0.29 }$}
&\textcolor{red}{$ 0.03 ^{+ 0.3 } _{ -0.33 }$}
&\textcolor{red}{$ 0.42 ^{+ 0.42 } _{ -0.32 }$}
&\textcolor{red}{{\bf 40}} &\textcolor{red}{$ 4.3 $} &\textcolor{red}{$ 24 $} &\textcolor{red}{$ 17 $} &\textcolor{red}{$ 2.0 $} \\
\hline%47
%\rowcolor{GRAY}
\textcolor{red}{GW190413\_134308} &\textcolor{red}{$ 51.37 ^{+ 23.14 } _{ -12.25 }$}
&\textcolor{red}{$ 33.63 ^{+ 11.68 } _{ -11.2 }$}
&\textcolor{red}{$ 35.97 ^{+ 7.43 } _{ -6.4 }$}
&\textcolor{red}{$ 0.67 ^{+ 0.29 } _{ -0.35 }$}
&\textcolor{red}{$ 0.02 ^{+ 0.27 } _{ -0.28 }$}
&\textcolor{red}{$ 0.61 ^{+ 0.32 } _{ -0.46 }$}
&\textcolor{red}{$ 5.5 $} &\textcolor{red}{{\bf 19}} &\textcolor{red}{$ 4.9 $} &\textcolor{red}{$ 13 $} &\textcolor{red}{$ 1.5 $} \\

\hline%13
GW190421\_213856
&$ 40.92 ^{+ 10.83 } _{ -6.62 }$
&$ 31.49 ^{+ 7.33 } _{ -8.08 }$
&$ 30.88 ^{+ 5.38 } _{ -3.85 }$
&$ 0.79 ^{+ 0.18 } _{ -0.3 }$
&$ -0.02 ^{+ 0.23 } _{ -0.27 }$
&$ 0.5 ^{+ 0.38 } _{ -0.37 }$
&$ 3.6 $ &\bf{11}&$ 1.3 $ &$ 6.1 $ &$ 1.7 $ \\
\hline%14
GW190424\_180648
&$ 40.22 ^{+ 11.78 } _{ -7.18 }$
&$ 31.11 ^{+ 7.46 } _{ -7.44 }$
&$ 30.57 ^{+ 5.67 } _{ -4.13 }$
&$ 0.79 ^{+ 0.18 } _{ -0.3 }$
&$ 0.17 ^{+ 0.23 } _{ -0.23 }$
&$ 0.5 ^{+ 0.38 } _{ -0.36 }$
&$ 2.8 $ &\bf{10.4}&$ 1.0 $ &$ 4.2 $ &$ 1.1 $ \\
\hline%15
%\rowcolor{GRAY}
\textcolor{red}{GW190426\_152155}
&\textcolor{red}{$ 5.18 ^{+ 1.92 } _{ -2.26 }$}
&\textcolor{red}{$ 1.55 ^{+ 1.04 } _{ -0.34 }$}
&\textcolor{red}{$ 2.39 ^{+ 0.08 } _{ -0.08 }$}
&\textcolor{red}{$ 0.3 ^{+ 0.59 } _{ -0.13 }$}
&\textcolor{red}{$ -0.11 ^{+ 0.22 } _{ -0.31 }$}
&\textcolor{red}{$ 0.18 ^{+ 0.39 } _{ -0.14 }$}
&\textcolor{red}{{\bf$>$10$^3$}} 
&\textcolor{red}{$ 606 $} 
&\textcolor{red}{{\bf$>$10}$^3$} 
&\textcolor{red}{$>${\bf10}$^3$} 
&\textcolor{red}{$>${\bf10}$^3$}
\\
\hline%16
GW190503\_185404
&$ 42.52 ^{+ 11.2 } _{ -7.73 }$
&$ 28.17 ^{+ 8.37 } _{ -8.89 }$
&$ 29.7 ^{+ 4.51 } _{ -4.08 }$
&$ 0.67 ^{+ 0.3 } _{ -0.29 }$
&$ -0.01 ^{+ 0.23 } _{ -0.25 }$
&$ 0.4 ^{+ 0.42 } _{ -0.3 }$
&\bf{5.0}&$ 4.6 $ &$ 2.6 $ &$ 3.8 $ &$ 1.5 $ \\
\hline%17
GW190512\_180714
&$ 22.97 ^{+ 5.44 } _{ -5.66 }$
&$ 12.5 ^{+ 3.52 } _{ -2.5 }$
&$ 14.48 ^{+ 1.31 } _{ -0.94 }$
&$ 0.54 ^{+ 0.37 } _{ -0.18 }$
&$ 0.03 ^{+ 0.12 } _{ -0.13 }$
&$ 0.23 ^{+ 0.37 } _{ -0.18 }$
&$ 0.4 $ &\bf{0.9}&$ 0.6 $ &$ 0.3 $ &$ 0.7 $ \\
\hline%18
GW190513\_205416
&$ 35.96 ^{+ 9.57 } _{ -9.22 }$
&$ 17.84 ^{+ 7.36 } _{ -4.26 }$
&$ 21.51 ^{+ 3.68 } _{ -1.9 }$
&$ 0.5 ^{+ 0.41 } _{ -0.18 }$
&$ 0.13 ^{+ 0.3 } _{ -0.18 }$
&$ 0.3 ^{+ 0.35 } _{ -0.22 }$
&$ 1.3 $ &$ 1.5 $ &\bf{1.9}&$ 0.4 $ &$ 0.4 $ \\
\hline%19
%\rowcolor{GRAY}
%\rowfont{\color{red}}
\textcolor{red}{GW190514\_065416}
&\textcolor{red}{$ 38.77 ^{+ 13.72 } _{ -7.96 }$}
&\textcolor{red}{$ 28.81 ^{+ 8.13 } _{ -8.26 }$}
&\textcolor{red}{$ 28.66 ^{+ 6.51 } _{ -4.55 }$}
&\textcolor{red}{$ 0.76 ^{+ 0.21 } _{ -0.32 }$}
&\textcolor{red}{$ -0.1 ^{+ 0.29 } _{ -0.35 }$}
&\textcolor{red}{$ 0.52 ^{+ 0.37 } _{ -0.36 }$}
&\textcolor{red}{$ 6.9 $} &\textcolor{red}{\bf{8.4}} &\textcolor{red}{$ 1.1 $} &\textcolor{red}{$ 12 $} &\textcolor{red}{$ 3.1 $} \\
\hline%20
%\rowcolor{GRAY}
\textcolor{red}{GW190517\_055101}
&\textcolor{red}{$ 36.7 ^{+ 11.44 } _{ -7.61 }$}
&\textcolor{red}{$ 25.49 ^{+ 6.68 } _{ -7.11 }$}
&\textcolor{red}{$ 26.51 ^{+ 4.0 } _{ -4.01 }$}
&\textcolor{red}{$ 0.7 ^{+ 0.26 } _{ -0.29 }$}
&\textcolor{red}{$ 0.56 ^{+ 0.19 } _{ -0.18 }$}
&\textcolor{red}{$ 0.46 ^{+ 0.29 } _{ -0.27 }$}
&\textcolor{red}{\bf{88}} &\textcolor{red}{$ 4.4 $} &\textcolor{red}{$ 1.1 $} &\textcolor{red}{$ 77 $} &\textcolor{red}{$ 9.1 $} \\
\hline%21
%\rowcolor{GRAY}
\textcolor{red}{GW190519\_153544}
&\textcolor{red}{$ 65.75 ^{+ 10.78 } _{ -11.27 }$}
&\textcolor{red}{$ 40.28 ^{+ 10.41 } _{ -10.14 }$}
&\textcolor{red}{$ 44.23 ^{+ 6.3 } _{ -6.15 }$}
&\textcolor{red}{$ 0.61 ^{+ 0.26 } _{ -0.18 }$}
&\textcolor{red}{$ 0.35 ^{+ 0.19 } _{ -0.24 }$}
&\textcolor{red}{$ 0.46 ^{+ 0.33 } _{ -0.3 }$}
&\textcolor{red}{\bf{34}} &\textcolor{red}{$ 24 $} &\textcolor{red}{$ 15 $} &\textcolor{red}{$ 28 $} &\textcolor{red}{$ 2.6 $} \\
\hline%22
GW190521\_074359
&$ 42.19 ^{+ 5.9 } _{ -4.91 }$
&$ 32.5 ^{+ 5.33 } _{ -5.77 }$
&$ 31.91 ^{+ 3.07 } _{ -2.17 }$
&$ 0.77 ^{+ 0.2 } _{ -0.2 }$
&$ 0.11 ^{+ 0.09 } _{ -0.14 }$
&$ 0.42 ^{+ 0.31 } _{ -0.3 }$
&\bf{1.8}&$ 6.0 $ &$ 1.7 $ &$ 0.9 $ &$ 0.6 $ \\
\hline%23
%\rowcolor{GRAY}
\textcolor{red}{GW190521}
&\textcolor{red}{$ 98.9 ^{+ 42.08 } _{ -18.79 }$}
&\textcolor{red}{$ 71.13 ^{+ 21.01 } _{ -27.9 }$}
&\textcolor{red}{$ 71.3 ^{+ 15.01 } _{ -9.92 }$}
&\textcolor{red}{$ 0.74 ^{+ 0.23 } _{ -0.42 }$}
&\textcolor{red}{$ 0.06 ^{+ 0.34 } _{ -0.35 }$}
&\textcolor{red}{$ 0.74 ^{+ 0.21 } _{ -0.4 }$}
&\textcolor{red}{$ 64 $} &\textcolor{red}{\bf{524}} &\textcolor{red}{$ 44 $} &\textcolor{red}{$ 409 $} &\textcolor{red}{$ 4.5 $} \\
\hline%24
GW190527\_092055
&$ 37.5 ^{+ 19.44 } _{ -10.13 }$
&$ 21.53 ^{+ 9.54 } _{ -8.0 }$
&$ 24.14 ^{+ 7.29 } _{ -4.0 }$
&$ 0.59 ^{+ 0.35 } _{ -0.32 }$
&$ 0.18 ^{+ 0.27 } _{ -0.29 }$
&$ 0.46 ^{+ 0.39 } _{ -0.35 }$
&\bf{1.9}&$ 3.6 $ &$ 1.7 $ &$ 1.4 $ &$ 1.1 $ \\
\hline%25
GW190602\_175927
&$ 67.85 ^{+ 16.23 } _{ -12.49 }$
&$ 47.74 ^{+ 13.16 } _{ -16.89 }$
&$ 48.66 ^{+ 8.34 } _{ -8.11 }$
&$ 0.72 ^{+ 0.25 } _{ -0.32 }$
&$ 0.11 ^{+ 0.26 } _{ -0.26 }$
&$ 0.41 ^{+ 0.41 } _{ -0.31 }$
&$ 11.4 $ &\bf{20.8}&$ 9.4 $ &$ 7.6 $ &$ 1.1 $ \\
\hline%26
%\rowcolor{GRAY}
\textcolor{red}{GW190620\_030421}
&\textcolor{red}{$ 55.96 ^{+ 16.1 } _{ -11.71 }$}
&\textcolor{red}{$ 36.79 ^{+ 10.97 } _{ -12.24 }$}
&\textcolor{red}{$ 38.63 ^{+ 7.35 } _{ -5.84 }$}
&\textcolor{red}{$ 0.66 ^{+ 0.3 } _{ -0.29 }$}
&\textcolor{red}{$ 0.37 ^{+ 0.21 } _{ -0.24 }$}
&\textcolor{red}{$ 0.44 ^{+ 0.35 } _{ -0.28 }$}
&\textcolor{red}{\bf{19}}&\textcolor{red}{$ 12 $} &\textcolor{red}{$ 6.7 $} &\textcolor{red}{$ 13 $} &\textcolor{red}{$ 1.7 $} \\
\hline%27
GW190630\_185205
&$ 35.08 ^{+ 6.89 } _{ -5.9 }$
&$ 23.34 ^{+ 5.22 } _{ -4.84 }$
&$ 24.68 ^{+ 2.17 } _{ -1.93 }$
&$ 0.66 ^{+ 0.28 } _{ -0.21 }$
&$ 0.11 ^{+ 0.13 } _{ -0.13 }$
&$ 0.32 ^{+ 0.3 } _{ -0.23 }$
&$ 1.0 $ &\bf{1.7}&$ 1.0 $ &$ 0.3 $ &$ 0.3 $ \\
\hline%28
%\rowcolor{GRAY}
\textcolor{red}{GW190701\_203306}
&\textcolor{red}{$ 54.08 ^{+ 12.12 } _{ -7.95 }$}
&\textcolor{red}{$ 41.3 ^{+ 8.28 } _{ -11.68 }$}
&\textcolor{red}{$ 40.62 ^{+ 5.32 } _{ -4.83 }$}
&\textcolor{red}{$ 0.77 ^{+ 0.2 } _{ -0.31 }$}
&\textcolor{red}{$ -0.04 ^{+ 0.23 } _{ -0.3 }$}
&\textcolor{red}{$ 0.45 ^{+ 0.4 } _{ -0.34 }$}
&\textcolor{red}{$ 17 $} &\textcolor{red}{\bf{24}} &\textcolor{red}{$ 5.1 $} &\textcolor{red}{$ 19 $} &\textcolor{red}{$ 2.8 $} \\
\hline%29
%\rowcolor{GRAY}
\textcolor{red}{GW190706\_222641}
&\textcolor{red}{$ 66.31 ^{+ 13.98 } _{ -15.06 }$}
&\textcolor{red}{$ 38.01 ^{+ 12.88 } _{ -11.71 }$}
&\textcolor{red}{$ 42.68 ^{+ 8.02 } _{ -6.26 }$}
&\textcolor{red}{$ 0.58 ^{+ 0.33 } _{ -0.22 }$}
&\textcolor{red}{$ 0.32 ^{+ 0.25 } _{ -0.3 }$}
&\textcolor{red}{$ 0.39 ^{+ 0.37 } _{ -0.27 }$}
&\textcolor{red}{\bf{21}} &\textcolor{red}{$ 13 $} &\textcolor{red}{$ 18 $} &\textcolor{red}{$ 14 $} &\textcolor{red}{$ 1.6 $} \\
\hline%30
GW190708\_232457
&$ 17.48 ^{+ 4.73 } _{ -2.27 }$
&$ 13.08 ^{+ 2.02 } _{ -2.7 }$
&$ 13.09 ^{+ 0.88 } _{ -0.62 }$
&$ 0.75 ^{+ 0.21 } _{ -0.28 }$
&$ 0.02 ^{+ 0.1 } _{ -0.08 }$
&$ 0.3 ^{+ 0.43 } _{ -0.24 }$
&$ 0.4 $ &\bf{1.0}&$ 0.4 $ &$ 0.5 $ &$ 1.4 $ \\
\hline%31
GW190719\_215514
&$ 36.04 ^{+ 18.7 } _{ -10.3 }$
&$ 20.53 ^{+ 8.43 } _{ -6.89 }$
&$ 23.15 ^{+ 6.08 } _{ -3.69 }$
&$ 0.58 ^{+ 0.37 } _{ -0.3 }$
&$ 0.38 ^{+ 0.27 } _{ -0.33 }$
&$ 0.44 ^{+ 0.33 } _{ -0.3 }$
&\bf{3.7}&$ 3.1 $ &$ 1.5 $ &$ 2.6 $ &$ 1.2 $ \\
\hline%32
GW190720\_000836
&$ 12.67 ^{+ 4.7 } _{ -2.65 }$
&$ 7.84 ^{+ 1.94 } _{ -1.94 }$
&$ 8.64 ^{+ 0.67 } _{ -0.68 }$
&$ 0.62 ^{+ 0.32 } _{ -0.28 }$
&$ 0.18 ^{+ 0.13 } _{ -0.11 }$
&$ 0.29 ^{+ 0.36 } _{ -0.2 }$
&\bf{0.9}&$ 0.4 $ &$ 0.1 $ &$ 0.4 $ &$ 0.6 $ \\
\hline%33
GW190727\_060333
&$ 40.56 ^{+ 13.66 } _{ -6.77 }$
&$ 29.42 ^{+ 6.5 } _{ -8.04 }$
&$ 29.79 ^{+ 4.22 } _{ -3.45 }$
&$ 0.73 ^{+ 0.24 } _{ -0.32 }$
&$ 0.14 ^{+ 0.26 } _{ -0.22 }$
&$ 0.46 ^{+ 0.41 } _{ -0.35 }$
&$ 2.7 $ &\bf{6.5}&$ 1.9 $ &$ 3.2 $ &$ 0.9 $ \\
\hline%34
GW190728\_064510
&$ 12.06 ^{+ 4.53 } _{ -1.97 }$
&$ 8.2 ^{+ 1.55 } _{ -1.99 }$
&$ 8.59 ^{+ 0.54 } _{ -0.33 }$
&$ 0.68 ^{+ 0.27 } _{ -0.3 }$
&$ 0.12 ^{+ 0.13 } _{ -0.06 }$
&$ 0.28 ^{+ 0.33 } _{ -0.2 }$
&\bf{1.0}&$ 0.4 $ &$ 0.1 $ &$ 0.3 $ &$ 0.5 $ \\
\hline%35
GW190731\_140936
&$ 41.98 ^{+ 14.82 } _{ -9.13 }$
&$ 29.1 ^{+ 9.33 } _{ -9.12 }$
&$ 30.02 ^{+ 6.83 } _{ -5.15 }$
&$ 0.71 ^{+ 0.26 } _{ -0.32 }$
&$ 0.17 ^{+ 0.3 } _{ -0.29 }$
&$ 0.5 ^{+ 0.36 } _{ -0.37 }$
&$ 3.4 $ &\bf{6.3}&$ 1.7 $ &$ 3.5 $ &$ 1.1 $ \\
\hline%36
GW190803\_022701
&$ 37.76 ^{+ 12.35 } _{ -7.33 }$
&$ 27.13 ^{+ 7.45 } _{ -7.69 }$
&$ 27.58 ^{+ 5.28 } _{ -3.99 }$
&$ 0.73 ^{+ 0.24 } _{ -0.3 }$
&$ 0.03 ^{+ 0.26 } _{ -0.27 }$
&$ 0.43 ^{+ 0.42 } _{ -0.33 }$
&$ 3.8 $ &$ 4.6 $ &$ 1.2 $ &\bf{6.3}&$ 1.7 $ \\
\hline%37
GW190814
&$ 23.21 ^{+ 1.02 } _{ -0.9 }$
&$ 2.59 ^{+ 0.08 } _{ -0.08 }$
&$ 6.1 ^{+ 0.06 } _{ -0.05 }$
&$ 0.11 ^{+ 0.01 } _{ -0.01 }$
&$ 0.0 ^{+ 0.06 } _{ -0.06 }$
&$ 0.04 ^{+ 0.04 } _{ -0.03 }$
&$ 0.1 $ &$ 0.4 $ &\bf{36}&$ 2.1 $ &$ 0.5 $ \\
\hline%38
GW190828\_063405
&$ 31.9 ^{+ 5.85 } _{ -3.77 }$
&$ 26.05 ^{+ 4.25 } _{ -4.6 }$
&$ 24.88 ^{+ 3.11 } _{ -1.92 }$
&$ 0.83 ^{+ 0.15 } _{ -0.23 }$
&$ 0.22 ^{+ 0.14 } _{ -0.16 }$
&$ 0.46 ^{+ 0.35 } _{ -0.32 }$
&$ 2.7 $ &\bf{4.3}&$ 0.5 $ &$ 2.2 $ &$ 1.0 $ \\
\hline%39
GW190828\_065509
&$ 23.73 ^{+ 7.22 } _{ -7.08 }$
&$ 10.18 ^{+ 3.55 } _{ -2.12 }$
&$ 13.22 ^{+ 1.18 } _{ -0.93 }$
&$ 0.43 ^{+ 0.39 } _{ -0.16 }$
&$ 0.08 ^{+ 0.16 } _{ -0.16 }$
&$ 0.31 ^{+ 0.38 } _{ -0.23 }$
&$ 0.6 $ &\bf{0.8}&$ 0.7 $ &$ 0.5 $ &$ 1.7 $ \\
\hline%40
%\rowcolor{GRAY}
\textcolor{red}{GW190909\_114149}
&\textcolor{red}{$ 43.09 ^{+ 50.64 } _{ -12.09 }$}
&\textcolor{red}{$ 27.68 ^{+ 12.7 } _{ -10.67 }$}
&\textcolor{red}{$ 29.48 ^{+ 17.33 } _{ -6.3 }$}
&\textcolor{red}{$ 0.63 ^{+ 0.32 } _{ -0.38 }$}
&\textcolor{red}{$ -0.03 ^{+ 0.46 } _{ -0.36 }$}
&\textcolor{red}{$ 0.53 ^{+ 0.38 } _{ -0.39 }$}
&\textcolor{red}{$ 6.0 $} &\textcolor{red}{$ 10 $} &\textcolor{red}{$ 2.1 $} &\textcolor{red}{\bf{15}} &\textcolor{red}{$ 4.5 $} \\
\hline%41
GW190910\_112807
&$ 43.28 ^{+ 7.59 } _{ -6.23 }$
&$ 34.98 ^{+ 6.34 } _{ -6.99 }$
&$ 33.69 ^{+ 4.32 } _{ -3.86 }$
&$ 0.82 ^{+ 0.15 } _{ -0.23 }$
&$ 0.03 ^{+ 0.18 } _{ -0.18 }$
&$ 0.41 ^{+ 0.39 } _{ -0.32 }$
&$ 3.3 $ &\bf{8.1}&$ 1.8 $ &$ 2.3 $ &$ 0.9 $ \\
\hline%42
GW190915\_235702
&$ 32.74 ^{+ 8.29 } _{ -4.72 }$
&$ 25.29 ^{+ 4.6 } _{ -5.57 }$
&$ 24.86 ^{+ 3.01 } _{ -2.35 }$
&$ 0.78 ^{+ 0.19 } _{ -0.29 }$
&$ 0.02 ^{+ 0.18 } _{ -0.22 }$
&$ 0.55 ^{+ 0.35 } _{ -0.4 }$
&$ 1.3 $ &\bf{8.0}&$ 0.6 $ &$ 2.9 $ &$ 1.4 $ \\
\hline%43
GW190924\_021846
&$ 8.59 ^{+ 2.85 } _{ -1.77 }$
&$ 5.14 ^{+ 1.21 } _{ -1.11 }$
&$ 5.74 ^{+ 0.26 } _{ -0.21 }$
&$ 0.6 ^{+ 0.33 } _{ -0.25 }$
&$ 0.02 ^{+ 0.14 } _{ -0.08 }$
&$ 0.21 ^{+ 0.34 } _{ -0.16 }$
&$ 0.1 $ &$ 0.2 $ &$ 0.05 $ &\bf{7.5}&$ 2.2 $ \\
\hline%44
GW190929\_012149
&$ 64.38 ^{+ 21.8 } _{ -19.12 }$
&$ 25.77 ^{+ 14.29 } _{ -9.6 }$
&$ 34.26 ^{+ 8.57 } _{ -6.33 }$
&$ 0.4 ^{+ 0.41 } _{ -0.19 }$
&$ 0.04 ^{+ 0.27 } _{ -0.27 }$
&$ 0.41 ^{+ 0.41 } _{ -0.31 }$
&$ 4.8 $ &$ 5.7 $ &\bf{49}&$ 2.9 $ &$ 1.3 $ \\
\hline%45
GW190930\_133541
&$ 12.05 ^{+ 6.44 } _{ -2.14 }$
&$ 7.9 ^{+ 1.62 } _{ -2.41 }$
&$ 8.45 ^{+ 0.5 } _{ -0.45 }$
&$ 0.66 ^{+ 0.28 } _{ -0.36 }$
&$ 0.14 ^{+ 0.2 } _{ -0.13 }$
&$ 0.32 ^{+ 0.38 } _{ -0.23 }$
&\bf{0.8}&$ 0.4 $ &$ 0.1 $ &$ 0.3 $ &$ 0.6 $ \\
\hline%46
%GW190803 &$ 37.76 ^{+ 12.35 } _{ -7.33 }$
%&$ 27.13 ^{+ 7.45 } _{ -7.69 }$
%&$ 27.58 ^{+ 5.28 } _{ -3.99 }$
%&$ 0.73 ^{+ 0.24 } _{ -0.3 }$
%&$ 0.03 ^{+ 0.26 } _{ -0.27 }$
%&$ 0.43 ^{+ 0.42 } _{ -0.33 }$
%&$ 3.7 $ &$ 4.6 $ &$ 1.2 $ &$ 6.0 $ &$ 1.6 $ \\

%\hline
%\hline
\end{tabular}
\end{center}
\caption{{\bf GWTC-2 gravitational-wave event parameters and Bayes factors.} 
Columns 2-7 show the median and 90\% symmetric credible intervals on selected source parameters: primary black hole mass $m_1$, secondary black hole mass $m_2$, chirp mass $\cal{M}$, mass ratio $q$, effective spin $\chi_{\rm eff}$ and precessing spin $\chi_{\rm p}$, taken from \cite{2020arXiv201014527A}. Columns 8-11 report our estimated Bayes factors for AGN vs GWTC-2 for different subsets of parameters. Column 12 reports our estimated Bayes factor for Tagawa-21 vs GWTC-2 on $\cal{M}$-${\chi_{\rm eff}}$-${\chi_{\rm p}}$ parameter space. The rows in red indicate that the event has a Bayes factor $>10$ 
for the $\cal{M}$-${\chi_{\rm eff}}$-${\chi_{\rm p}}$ parameter space. For each row we highlighted the highest Bayes factor.} 
\end{table*}

\begin{acknowledgments}
%The authors are grateful to XY for useful suggestions.
We gratefully acknowledge the support of LIGO and Virgo for provision of computational resources.
I.B. acknowledges the support of the National Science Foundation under grant \#1911796 and of the Alfred P. Sloan Foundation. ZH was supported by NASA grant NNX15AB19G and NSF grants AST-2006176 and AST-1715661.
This research has made use of data, software and/or web tools obtained from the Gravitational Wave Open Science Center, a service of LIGO Laboratory, the LIGO Scientific Collaboration and the Virgo Collaboration. LIGO is funded by the U.S. National Science Foundation. Virgo is funded by the French Centre National de Recherche Scientifique (CNRS), the Italian Istituto Nazionale della Fisica Nucleare (INFN) and the Dutch Nikhef, with contributions by Polish and Hungarian institutes. HT acknowledges support by the Grants-in-Aid for Basic Research by the Ministry of Education, Science and Culture of Japan (HT:17H01102, 17H06360).  This material is based upon work supported by NSF’s LIGO Laboratory which is a major facility fully funded by the National Science Foundation.
\end{acknowledgments}

\bibliography{reference}

\end{document}